\documentclass[final]{aipproc}

\layoutstyle{6x9}

\begin{document}

\title{The CMS Silicon Strip Tracker}


%{LHC, CMS, central tracker, silicon strip tracker, 
%sensors, hybrids, modules,
%read out electronics, control electronics, mass production, integration, 
%test beam}

\author{Gabriella P\'asztor \\ {\small for the CMS Collaboration}}
{
  address={Department of Physics, University of California, 
  Riverside, CA92521, USA},
  altaddress={KFKI RMKI, 
  Konkoly Thege Mikl\'os \'ut 29-33, Budapest, H-1121, Hungary} 
  % additional visiting address
}

\begin{abstract}
The CMS collaboration is constructing the largest silicon tracker ever built
with an active silicon area of 200 m$^2$ to provide robust charged
particle tracking and vertex reconstruction within the 4T magnetic field 
of the CMS Solenoid.  
The design of the detector, the status of the construction and the performance
of the substructures are reviewed.
\end{abstract}

\maketitle

\section{The CMS Silicon Strip Tracker}

The CMS Tracker is composed of a Silicon Pixel Detector and a large volume
Silicon Strip Tracker (SST). The CMS SST~\cite{cms-sst} consists of 15148
modules housing 24244 silicon strip sensors and their front-end electronics
with 9.6 million readout channels in total. The modules are mounted on local
support structures in four distinct subsystems: the Inner and Outer Barrels,
the two Inner Disks and the two End-Caps.  The Inner Barrel is composed of four
and the Outer Barrel of six cylindrical layers. Each Inner Disk is made of three disks,
each divided into three rings,  while the End-Caps have nine disks with four to
seven rings. Typically 10 points are measured along the track up to a rapidity
of 2.4. The regions $20<r<40$ and $60<r<75$~cm are populated by double sided
modules constructed by mounting two independent single sided modules back to
back with a stereo angle of 100 mrad. The SST will be housed inside a 5.4 m
long and a 2.4 m diameter cylindrical support tube and an active thermal shield
will keep the volume at a temperature below -10 C and at a relative humidity of
30\%.

\section{Module and component production}

The basic building blocks of the SST are the modules. Each module has 1 or 2
silicon sensors mounted on a carbon-fiber or graphite support frame  with a
kapton circuit to isolate the silicon backplane and supply the bias voltage. 
The readout and control chips are mounted on the front-end (FE) hybrid. A glass
circuit, the pitch adapter, provides fan-out from the sensors with a pitch of
80-205 $\mu$m to the readout chips. In total about 25M micro-bonds provide the
electrical connection between the module parts. To comply with the mechanical
constraints there are 29 different module types using 16 different single-sided
sensor, 12 FE hybrid and 26 pitch adapter designs.

Sensors~\cite{sensor} are fabricated on 6" wafers 
using  non-oxygenated n-type bulk with <100> lattice orientation and have 512
or 768 p$^+$ strip implants with a width/pitch ratio of 0.25. In the
barrel region, rectangular sensors are mounted with strips parallel to the beam
axis, while in the End-Caps and Inner Disks, wedge-shaped sensors 
with strips in the radial direction. In the inner region ($r<60$~cm) one 
320 $\mu$m thick, low resistivity (1.5-3.5 k$\Omega$cm) sensor is housed
on a module, while in  the outer region two 500 $\mu$m thick, high
resistivity (3.5-7.5 k$\Omega$cm) sensors are daisy-chained. 

During the production of the thick sensors a large number of problems 
were encountered, most notably
a deep corrosion of the sensors~\cite{sensor-corrosion}. This forced us to 
shift most of the order to the producer of the thin sensors.

The front-end (FE) hybrid is realized as a four-layer kapton substrate
laminated on a ceramic carrier and having a flex tail for electrical 
connections. It houses 4 or 6 APV25 readout chips made by IBM's radiation hard
0.25 $\mu$m CMOS technology. 
% The APV25 has a charge sensitive amplifier with
% $\tau$=50 ns, a CR-RC shaper and 192 pipeline cells (4.8 $\mu$s) per channel,
% and it multiplexes 128 channels to one analog output.
The APV25 has two operation
modes: the peak mode where the output sample corresponds to the peak amplitude, 
and the deconvolution mode that makes a weighted sum of three consecutive 
samples and
allows for the identification of the correct bunch crossing in the  high
luminosity phase of LHC. 

Several problems occurred during the FE hybrid mass
production, the most serious of which was the discovery of broken vias which
turned out to be inherent to the design and the production method. The FE 
hybrid originally was built up from two kapton circuits glued together. During
the laser drilling of the 100 $\mu$m vias, the glue melted faster, potentially
forming a cavity which then lead to very poor metalization of the via. This
problem was solved by introducing a third kapton layer and by 
increasing the via diameter from 100 to 120 $\mu$m. Throughout the production,
the long-term reliability of the micro-bonds between the APV25 and
the substrate due to over-deformation of the bond feet and the occurrence of
cratering was also a concern.

All modules are built and tested  in the collaborating institutes with
automatized, computer-controlled procedures, including a high-precision robotic
assembly of the modules at six production centers and  wire bonding using more
than 20 bonding machines. The modules are tested on fast single
module setups based on the APV Readout Controller (ARC) and then on long-term
test stands with a CMS-like DAQ system executing several thermal cycles. 
Module mass production started in 2004 and after several production stops, it
is expected to finish in early 2006. The production yield varies between 99 and
94\% depending on the subsystem, with an excellent rate of typically $0.1-0.3$\%
bad strips per module.

\section{Integration and Subsystem Performance}

The modules are mounted onto light modular carbon-fiber substructures  which
also house the printed circuit boards for control electronics and for  optical
signal transmission from the detector to the surface electronics building. The
light cooling pipes are an integral part of the supporting mechanics but not
structural components. The modules are fixed to supporting blocks made from
aluminum or composite material. The blocks are machined with a precision better
than 20 $\mu$m and are in direct dry contact with the cooling pipes. The final
module mounting precision within a subsystem is expected to be better than 200
$\mu$m. The integration of modules into the large substructures are under way 
as shown in Figure 1.

\begin{figure}

\begin{minipage}{8cm}
\includegraphics[width=.25\textheight, bb=57 91 542 757,angle=-90]{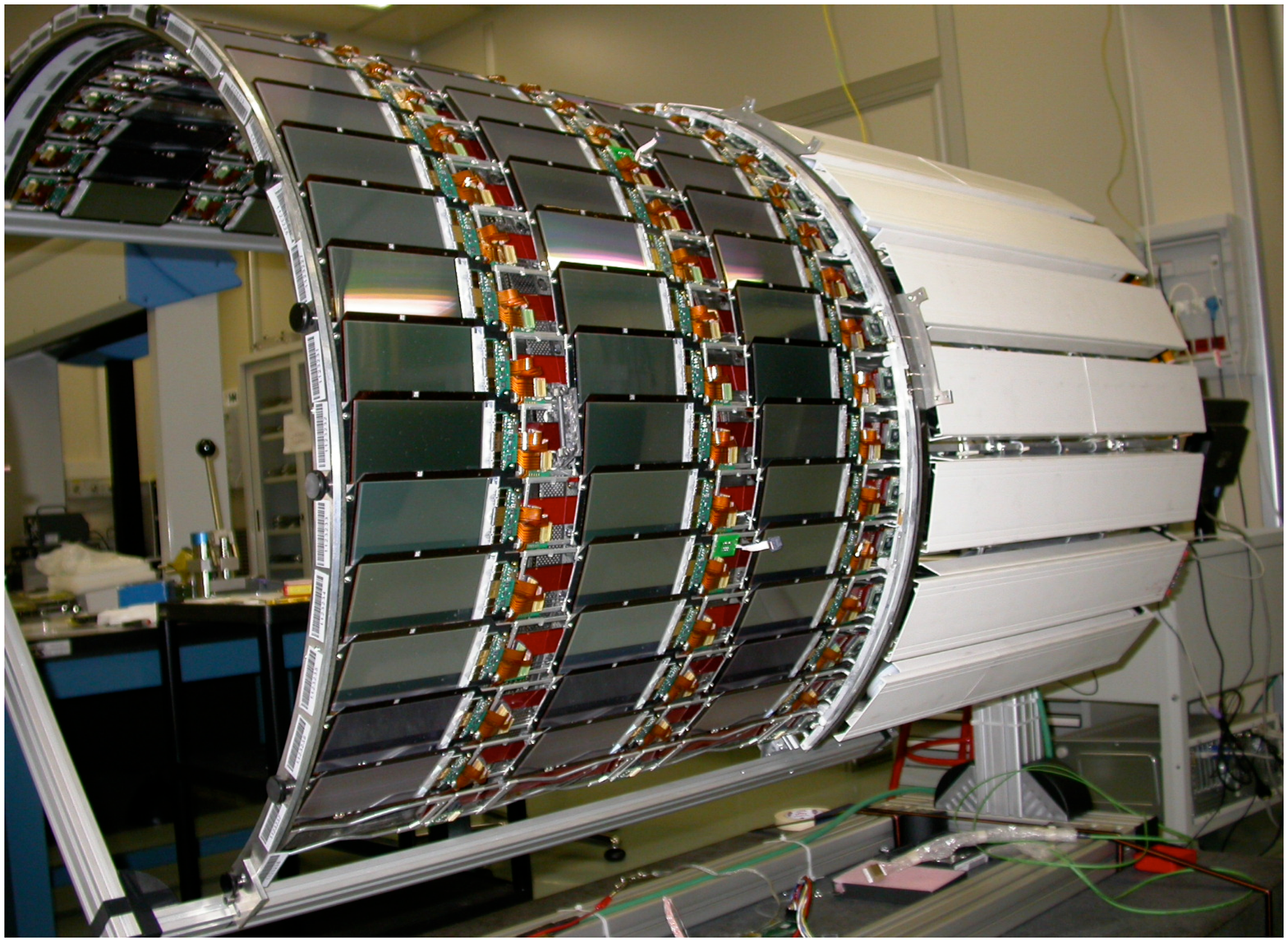}
\end{minipage}

\begin{minipage}{6cm}
\includegraphics*[height=.25\textheight, bb=90 200 575 665]{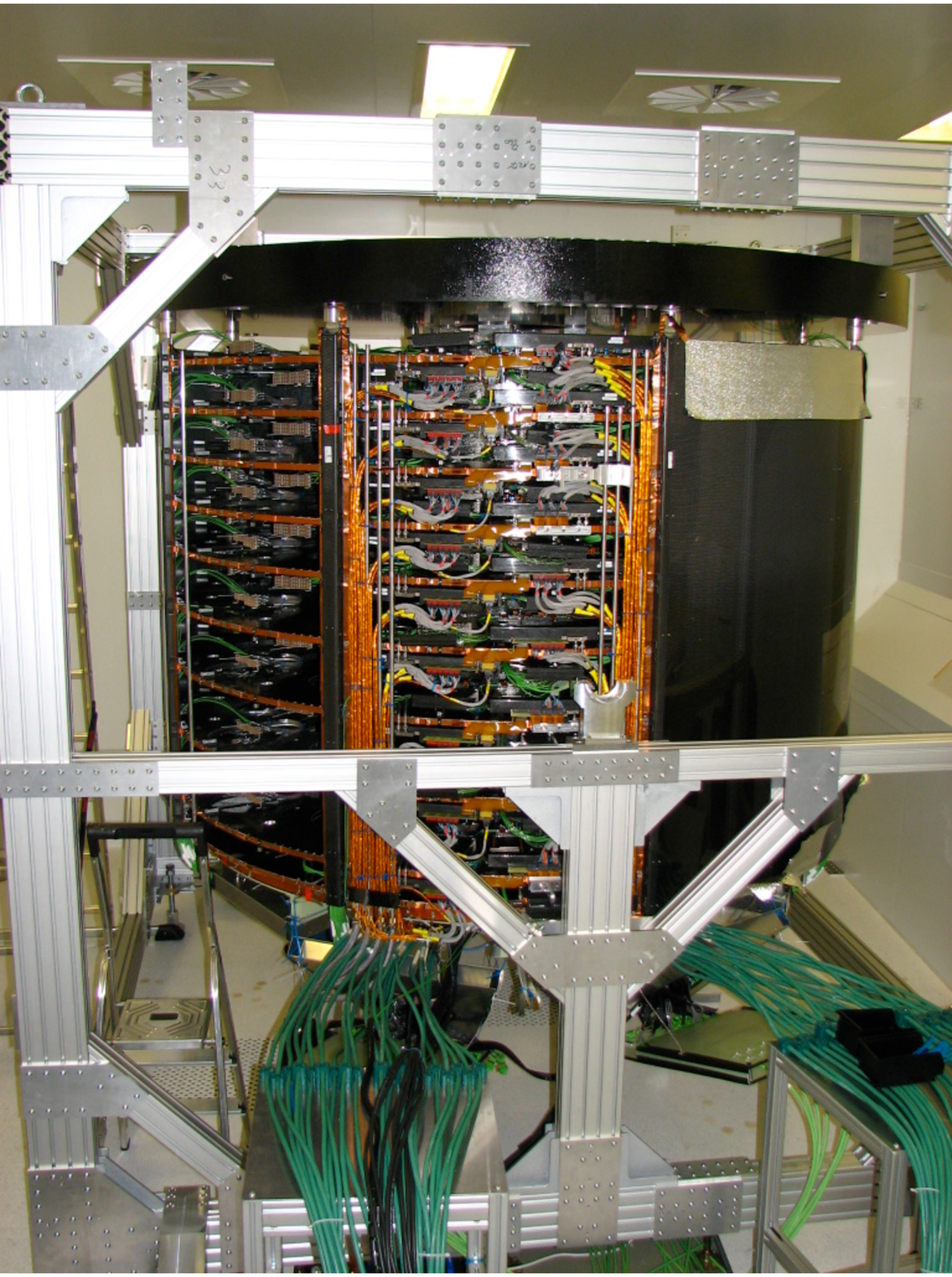}
\end{minipage}

\caption{
(Left) One of the four carbon-fiber half-shells of layer 3 of the Inner Barrel.
%(Center) The Outer Barrel structure partially inserted into the Tracker 
% support tube.
(Right) The first integrated sector of one of the End-Caps.}
\end{figure}

The performance of the substructures was measured in test beam experiments and
system tests. In 2004 around 1\% of all three subsystems were tested at CERN
X5 with 120 GeV pions and $70-120$~GeV muons both at room temperature and at
around $-10$~C. We have experienced stable communication and readout, uniform
noise distribution with small common mode noise. The signal/noise (S/N) ratio
was found to be around 20 and the equivalent noise charge (ENC) consistent with
expectations. The results~\cite{testbeam} are shown on Figure 2.
  
\begin{figure}
\includegraphics[height=.22\textheight]{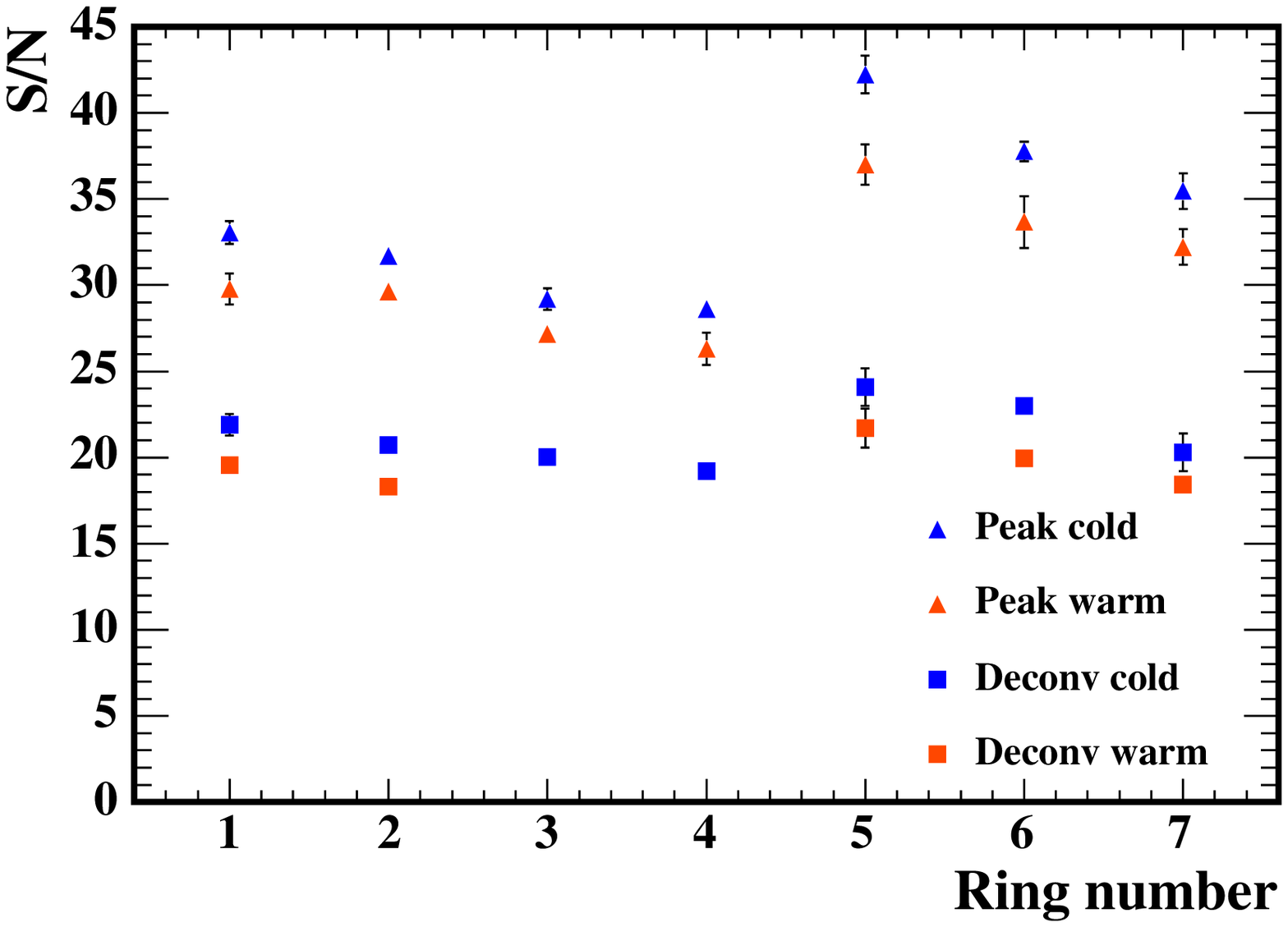}
\includegraphics[height=.22\textheight]{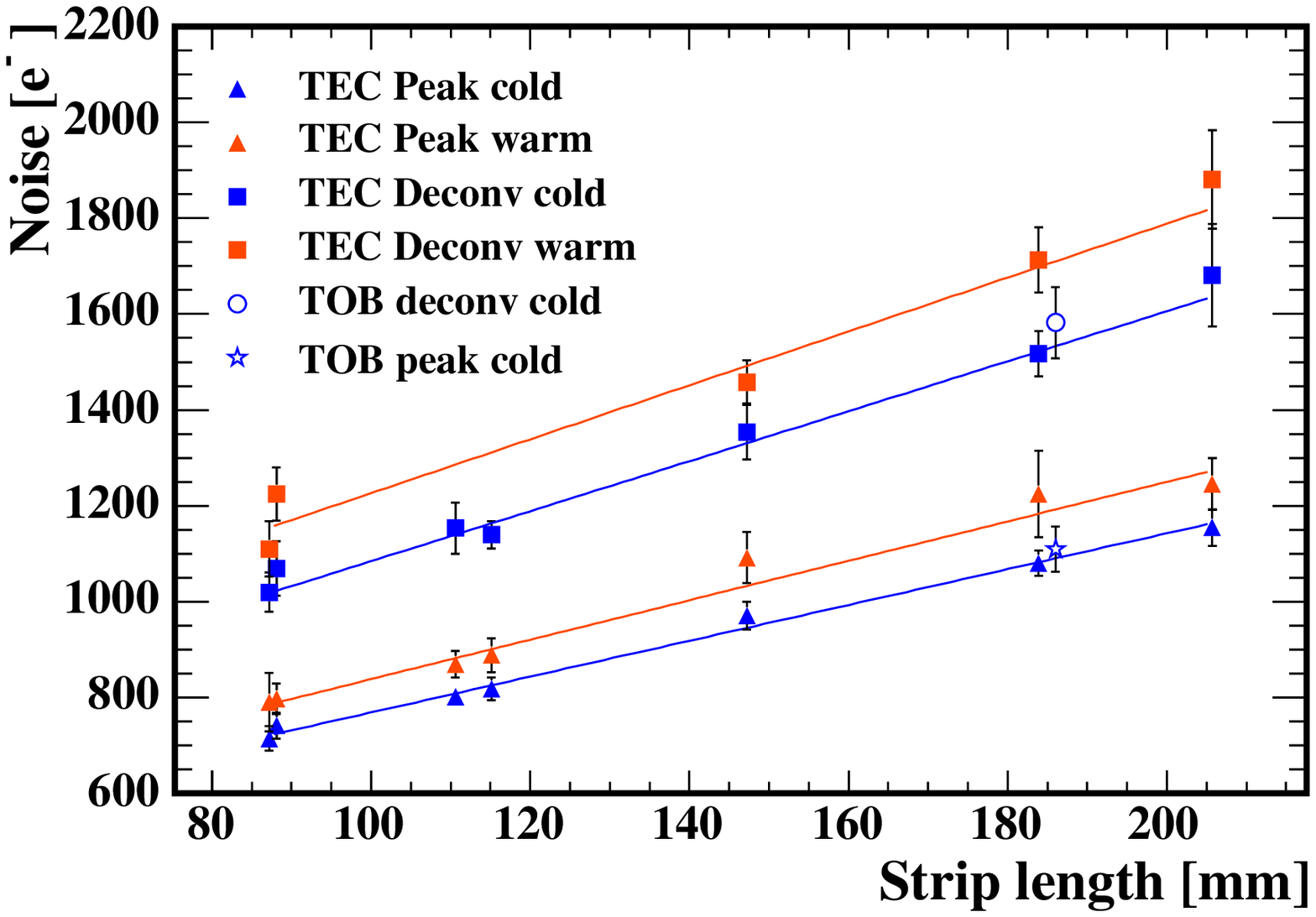}
\caption{(Left) Mean S/N for each ring of the End-Cap system.
(Right) The mean ENC as a function of the strip length for the End-Cap (TEC)
and the Outer Barrel (TOB) systems.}
\end{figure}

\end{document}